# CORRELATION FUNCTIONS AND PRESSURE OF NON-ISOTHERMAL PLASMA

Emile S. Medvedev

Present address: Institute of Problems of Chemical Physics, Russian Academy of Sciences, 142432 Chernogolovka, Russia
Email: esmedved@orc.ru
WEB page: http://www.esmedvedev.orc.ru/

**Abstract**
Expressions for correlation functions of classical non-isothermal two-component plasma are derived. In the limiting case of $\Theta_e \gg \Theta_i$ strong correlations arise due to the existence of weakly damping waves (ionic sound), whose phase velocity is between the thermal velocities of particles, $u_e \gg \omega/k \gg u_i$. Owing to these correlations, the additional term in the expression for pressure due to the Coulomb interaction changes its sign as compared to the case of thermodynamic equilibrium where such waves do not exist.

This paper considers the effect of weakly damping waves on correlation functions of the non-isothermal plasma. The method used (the method of probe particle [1]) is not rigorous, however, it is simple and physically transparent. The rigorous treatment requires investigations of solutions of the equation for the binary correlation function. In the classical case, such an equation appropriate for studies correlations in non-equilibrium systems was derived by Bogolyubov [2]. Generalized to quantum systems by Klimontovich and Temko [3], this equation was solved by Silin [4]. In the Appendix we analyze the solution of Bogolyubov's equation.

The potential of the electric field generated by a charge of sort β moving in plasma obeys the Poisson equation,

$$\Delta\varphi(\mathbf{r}) = -4\pi e_\beta(\mathbf{r} - \mathbf{x}(t)) - 4\pi \sum_\alpha e_\alpha \delta n_{\alpha\beta}(\mathbf{r}, \mathbf{x}(t)), \tag{1}$$

where $\delta n_{\alpha\beta}(r, x(t))$ is a perturbation of the charge density of sort α at point $r$. The system of hydrodynamics equations for particles of sort α has the form

$$m_\alpha n_\alpha \frac{\partial \mathbf{v}_\alpha}{\partial t} = -\nabla p_\alpha - e_\alpha n_\alpha \nabla \varphi, \tag{2a}$$

$$\frac{\partial n_\alpha}{\partial t} + n_\alpha \nabla \mathbf{v}_\alpha = 0, \tag{2b}$$

$$\frac{d}{dt}\left(\frac{p_\alpha}{n_\alpha^a}\right) = 0; \quad p_\alpha = n_{\alpha 0}\Theta_\alpha. \tag{2c}$$

Here $n_{\alpha 0}$ and $\Theta_\alpha$ are the average density and temperature of particles of sort α, and $a$ is a constant of the order of unity. Linearizing system (2) and using the Fourier transform of eq (1), we obtain the following system of linear equations for functions $\delta n_{\alpha\beta}(k, x(t))$:

$$\frac{\partial^2}{\partial t^2}\delta n_{\alpha\beta} + a(ku_\alpha)^2 \delta n_{\alpha\beta} + \omega_\alpha^2 \left(\sum_\gamma z_{\alpha\gamma}\delta n_{\gamma\beta} + z_{\alpha\beta}e^{-ikx(t)}\right) = 0,$$

$$\omega_\alpha^2 = \frac{4\pi n_{\alpha 0}e_\alpha^2}{m_\alpha}; \quad u_\alpha^2 = \frac{\Theta_\alpha}{m_\alpha}; \quad z_{\alpha\beta} = \frac{e_\beta}{e_\alpha}. \tag{3}$$

Neglecting the back effect of the plasma particles on the movement of the probe particle, we will assume $x(t) = x_0 + \mathbf{v}t$. Then, the solution of eq (3) is given by the formula

$$\delta n_{\alpha\beta}(k, x(t)) = \frac{z_{\alpha\beta}}{\mathrm{E}(k\mathbf{v}, k)}\frac{\omega_\alpha^2}{(k\mathbf{v})^2 - a(ku_\alpha)^2}e^{-ikx(t)}, \tag{4}$$

where

$$\mathrm{E}(\omega, k) = 1 - \sum_\gamma \frac{\omega_\gamma^2}{\omega^2 - a(ku_\gamma)^2}. \tag{5}$$

Now let us calculate the correlation function of density. By definition,

$$n_{\alpha 0}n_{\gamma 0}g_{\alpha\gamma}(r, r') = \sum_\beta n_{\beta 0}\int dx d\mathbf{v}\, f_\beta(\mathbf{v})[n_\alpha(r) - n_{\alpha 0}][n_\gamma(r') - n_{\gamma 0}], \tag{6}$$

where $f_\beta(\mathbf{v})$ is the distribution function normalized to unity and

$$n_\alpha(r) = n_{\alpha 0} + \delta_{\alpha\beta}\delta(r - x(t)) + \delta n_{\alpha\beta}(r, x(t)). \tag{7}$$

Inserting (7) into (6) and using (4) and (5), after integration over $x$, we find that $g_{\alpha\gamma}(r, r')$ depends only on $r - r'$. For the Fourier transform, we obtain

$$n_\alpha n_\gamma g_{\alpha\gamma}(\mathbf{k}) = \sum_\beta n_\beta \int\limits_{-\infty}^{+\infty} du f_\beta(u) \left[ \delta_{\alpha\beta} + \frac{z_{\alpha\beta}}{k^2 \mathrm{E}(ku,\mathbf{k})} \cdot \frac{\omega_\alpha^2}{u^2 - au_\alpha^2} \right]$$
$$\times \left[ \delta_{\gamma\beta} + \frac{z_{\gamma\beta}}{k^2 \mathrm{E}(-ku,-\mathbf{k})} \cdot \frac{\omega_\gamma^2}{u^2 - au_\gamma^2} \right],$$
(8)

where integration is performed over the velocity components perpendicular to $\mathbf{k}$ and

$$f_\beta(u) = \int d\mathbf{v} f_\beta(\mathbf{v}) \delta\left(u - \frac{\mathbf{kv}}{k}\right);$$

subscript 0 of the density is omitted. In what follows, the value of constant $a$ is insignificant since the result weakly depends on it, therefore, we set $a = 1$. The integral in (8) diverges in points where $\mathrm{E}(ku,\mathbf{k})$ vanishes. An analysis similar to the above but based on the kinetic equation shows that taking into account the damping of the oscillations removes the divergence. We will study the effect of plasma oscillations on the correlation functions in the limiting case of $\Theta_e \gg \Theta_i$ with adding the correct imaginary part to $\mathrm{E}(ku,\mathbf{k})$ [1].

Solving the equation $\mathrm{E}(ku,\mathbf{k}) = 0$, we find two kinds of oscillations: the Langmuir oscillations with the frequency $\omega^2 = \omega_e^2 + k^2 u_e^2$ and the ionic-sound waves,

$$\omega^2 = \omega_i^2 \left[ 1 + (kL_e)^{-2} \right]^{-1} \quad \left( L_e^2 = \frac{\Theta_e}{4\pi n_e e^2} \right).$$

The Langmuir oscillations are damping weakly at $kL_e \ll 1$, but at such $k$ their phase velocity $\omega/k$ is much greater than the thermal velocity of electrons, $u_e$. Therefore, these oscillations are excited by an exponentially small number of particles, and hence, the corresponding contribution to (8) is negligible. The situation is different for the ionic sound. Its phase velocity obeys inequalities $u_e \gg \omega/k \gg u_i$. This means, from one hand, that the oscillations damping is small and, from the other hand, there is a large group of particles (i.e., all electrons) that excite this kind of oscillations. As we will see below, this has a strong impact on the correlations as compared to the case of statistical equilibrium.

Let, for simplicity, plasma consist of electrons and one sort of ions, so that $n_i = n_e \equiv n$. According to the above said, the major contribution to (8) is provided by the region $u_e > u > u_i$. Within this interval one has

$$\mathrm{E}(ku,\mathbf{k}) \equiv \mathrm{E}*(-ku,-\mathbf{k}) = 1 + (kL_e)^{-2} - (\omega_i/ku)^2 - i\Gamma(u),$$
(9)

where $\Gamma(u) = \sqrt{\pi}\left(u/u_e\right)(kL_e)^{-2}$. Since $\Gamma^2 \ll |E|^2$ in the entire range except for the point $u = u_0$ that corresponds to the ionic sound, we set $\Gamma(u) \equiv \Gamma(u_0)$. Consider, first, $\alpha = \gamma = i$. Then, we obtain from (8) and (9):

$$ng_{ii}(k) = \frac{1}{u_e}\sqrt{\frac{2}{\pi}}\int_{u_i}^{u_e} du \left(\frac{\omega_i}{ku}\right)^4 \left\{\left[1+(kL_e)^{-2}-\left(\frac{\omega_i}{ku}\right)^2\right]^2 + \pi(kL_e)^{-4}\left(\frac{u_0}{u_e}\right)^2\right\}^{-1}. \quad (10)$$

The integration can be extended to the entire positive semi-axis due to the integral convergence. After simple transformations, eq (10) takes the form

$$ng_{ii}(k) = \frac{1}{u_e}\sqrt{\frac{2}{\pi}}\int_0^\infty du \left\{\left[\left(\frac{u}{u_0}\right)^2 - 1\right]^2 + \frac{\pi}{\mu^2}\left(\frac{u_0}{u_i}\right)^6\right\}^{-1}; \quad \mu = \frac{u_e\Theta_e}{u_i\Theta_i}. \quad (11)$$

The final result is

$$ng_{ii}(k) = 1 + (kL_e)^2, \quad kL_i \ll 1. \quad (12)$$

When $kL_i \gg 1$, $u_0$ becomes less than $u_i$; ionic sound becomes strongly damping and its contribution to $g_{ii}(k)$ decreases as $(kL_i)^{-4}$. Similarly, the following expressions are obtained for two other correlation functions:

$$ng_{ee}(k) = \frac{1}{2}\left[1+(kL_e)^2\right]^{-1}, \quad ng_{ie}(k) = 1 \quad (kL_i \ll 1). \quad (13)$$

Remind that, in the case of equilibrium, $\Theta_\gamma \equiv \Theta$, the Debye formula applies,

$$g_{\alpha\beta}(k) = -\frac{v_{\alpha\beta}(k)}{\Theta}\left[1+(kL)^{-2}\right]^{-1}, \quad v_{\alpha\beta}(k) = \frac{4\pi e_\alpha e_\beta}{k^2}, \quad L^{-2} = \sum_\gamma \frac{4\pi n_\gamma e_\gamma^2}{\Theta_\gamma}. \quad (14)$$

Similar terms are also present in the corresponding exact expressions (see eqs (A14)-(A16) in Appendix), however, they are small at $kL_i \ll 1$ (more accurately, at $k \leq k_i$, $k_i L_i = (2|\ln \mu|)^{-1/2}$).

Comparing (14) with (12) and (13), we see that, in contrast to the Debye case, the correlations due to the waves do not obey the rule "similarly charged particles are mutually repulsing, dissimilarly charged one are attracting", i.e., attraction of similar particles occurs. This can be understood if we take into account that the density fluctuations in two closely spaced points *A* and *B* induced by a remote electron must be identical due to homogeneity of the system. For instance, if an increase in the electron density occurs at point *A*, the electron density at *B* also increases, which means inter-electron attraction[1]. Moreover, the nature of the waves responsible for perturbation is of no importance; e.g., considering the correlations due to Langmuir oscillations, we would obtain a similar result. In contrast, the sign of the correlation function for dissimilar charges essentially depends on the nature of the waves. In the ionic-sound wave, equation $\text{div}\mathbf{j} = 0$ is fulfilled, which means that the electron and ion currents are directed oppositely to each other, hence, the particles themselves move in one and the same direction. Therefore, an increase in the electron density at a given point is accompanied by an increase in the ion density, hence, attraction takes place. On the contrary, in the Langmuir oscillations, the currents have the same directions, which results in the change of sign.

---

[1] At thermal equilibrium, the direct Coulomb repulsion is significantly stronger than the above effect.

The other interesting feature of eqs (12) and (13) is that $g_{ii}(k)$ is much greater in absolute value than $g_{ie}(k)$ and $g_{ee}(k)$ in the most essential region of $L_e^{-1} \ll k \ll L_i^{-1}$. The electrons exciting the ionic sound move with the velocities greatly exceeding the phase velocity of the waves. The front of the emitted wave has the shape of cone with the cone angle on the order of $\omega/ku_e \ll 1$. Therefore, the electron is correlated only with those particles that are within this cone. On the other hand, the ion correlations do not subject to such limitations.

Let us find the additional term in pressure due to the Coulomb interaction. By the virial theorem, we have

$$\Delta P \equiv P - n\left(\Theta_e + \Theta_i\right) = \frac{1}{3}E_{\text{coul}}. \tag{15}$$

In eq (A17) for $E_{\text{coul}}$, we can retain only the term with $g_{ii}(k)$. After integration over $k$ up to $k_i$, we obtain

$$\Delta P = \frac{1}{36\pi^2}\Theta_e k_i^3. \tag{16}$$

which is identical to (A20). We see that the effective interaction between particles, which arises due to the existence of non-damping slow waves, results in a change of sign of $\Delta P$ as compared to the case of thermal equilibrium.

The author is grateful to A. A. Vedenov for the proposed problem and great help.

APPENDIX

The Bogolyubov equation for the first term of expansion of the binary correlation function in powers of $N_D^{-1}$ ($N_D$ is the number of particles in the Debye sphere) has the form [2] (see also [5])

$$n_\alpha n_\beta V_{\alpha\beta}(k) g_{\alpha\beta}(k,\mathbf{v}_\alpha,\mathbf{v}_\beta) = (k\mathbf{v}_\alpha - k\mathbf{v}_\beta - i\Delta)^{-1} \Big[ n_\alpha n_\beta f_\alpha(\mathbf{v}_\alpha) f_\beta(\mathbf{v}_\beta)$$
$$\times V_{\alpha\beta}^2(k) \left( \frac{k\mathbf{v}_\beta}{\Theta_\beta} - \frac{k\mathbf{v}_\alpha}{\Theta_\alpha} \right) - n_\alpha V_{\alpha\alpha}(k) f_\alpha(\mathbf{v}_\alpha) \frac{k\mathbf{v}_\alpha}{\Theta_\alpha} h_\beta(-k,\mathbf{v}_\beta) \quad , \quad (A1)$$
$$+ n_\beta V_{\beta\beta}(k) f_\beta(\mathbf{v}_\beta) \frac{k\mathbf{v}_\beta}{\Theta_\beta} h_\alpha(k,\mathbf{v}_\alpha) \Big], \quad \Delta \to 0$$

where

$$g_{\alpha\beta}(r,\mathbf{v}_\alpha,\mathbf{v}_\beta) = \int \frac{dk}{(2\pi)^3} e^{ikr} g_{\alpha\beta}(k,\mathbf{v}_\alpha,\mathbf{v}_\beta),$$
$$g_{\alpha\beta}(r,\mathbf{v}_\alpha,\mathbf{v}_\beta) \equiv F_{\alpha\beta}(r,\mathbf{v}_\alpha,\mathbf{v}_\beta) - 1, \quad (A2)$$
$$h_\alpha(k,\mathbf{v}_\alpha) = \sum_\beta n_\alpha n_\beta V_{\alpha\beta}(k) \int d\mathbf{v}_\beta g_{\alpha\beta}(k,\mathbf{v}_\alpha,\mathbf{v}_\beta),$$

$$V_{\alpha\beta}(k) = \frac{4\pi e_\alpha e_\beta}{k^2}, \quad f_\alpha(\mathbf{v}_\alpha) = \left( \frac{m_\alpha}{2\pi\Theta_\alpha} \right)^{3/2} \exp\left( -\frac{m_\alpha \mathbf{v}_\alpha^2}{2\Theta_\alpha} \right). \quad (A3)$$

Here $F_{\alpha\beta}(r,\mathbf{v}_\alpha,\mathbf{v}_\beta)$ is the binary correlation function. The method used in ref [4] for solving the corresponding quantum equation applies, without any modifications, to the present case as well. Therefore, we will just write the result,

$$g_{\alpha\beta}(k,\mathbf{v}_\alpha,\mathbf{v}_\beta) = q_{\alpha\beta}(k,\mathbf{v}_\alpha,\mathbf{v}_\beta) + q^*_{\beta\alpha}(k,\mathbf{v}_\beta,\mathbf{v}_\alpha), \quad (A4)$$

$$q_{\alpha\beta}(k,\mathbf{v}_\alpha,\mathbf{v}_\beta) = -\frac{V_{\alpha\beta} f_\alpha(\mathbf{v}_\alpha) f_\beta(\mathbf{v}_\beta)}{k\mathbf{v}_\alpha - k\mathbf{v}_\beta - i\Delta} \cdot \frac{k\mathbf{v}_\alpha}{\Theta_\alpha} \left[ \frac{1}{\varepsilon^+(k\mathbf{v}_\beta,k)} - \frac{k\mathbf{v}_\beta}{\Theta_\beta} I_\beta \right], \quad (A5)$$

$$I_\beta = \sum_\gamma n_\gamma V_{\gamma\gamma} \int \frac{d\mathbf{v}_\gamma f_\gamma(\mathbf{v}_\gamma)}{k\mathbf{v}_\beta - k\mathbf{v}_\gamma + i\Delta} \left[ \varepsilon^+(k\mathbf{v}_\gamma,k) \varepsilon^-(k\mathbf{v}_\gamma,k) \right]^{-1}, \quad (A6)$$

$$\varepsilon^\pm(\omega,k) \equiv \varepsilon(\omega \pm i\Delta, k),$$
$$\varepsilon(\omega,k) = 1 + \sum_\alpha n_\alpha V_{\alpha\alpha} \int \frac{d\mathbf{v}_\alpha}{\omega - k\mathbf{v}_\alpha} \left[ k \frac{\partial f_\alpha(\mathbf{v}_\alpha)}{m_\alpha \partial \mathbf{v}_\alpha} \right]. \quad (A7)$$

Let us rewrite $I_\beta$ using the following identity:

$$\frac{1}{\varepsilon^+(\omega,k)\varepsilon^-(\omega,k)} \equiv \frac{1}{2 \operatorname{Im} \varepsilon^-(\omega,k)} \left[ \frac{1}{\varepsilon^+(\omega,k)} - \frac{1}{\varepsilon^-(\omega,k)} \right]. \quad (A8)$$

Then (A6) takes the form

$$I_\beta = \frac{1}{2\pi i k} \int_{-\infty}^{+\infty} \frac{dx}{x - x_\beta - i\Delta} \cdot \frac{\varphi(x)}{x} \left[ \frac{1}{\varepsilon^+(kx,k)} - \frac{1}{\varepsilon^-(kx,k)} \right], \quad (A9)$$

$$\mathbf{k}\mathbf{v}_\beta \equiv kx_\beta, \quad \mathbf{k}\mathbf{v}_\gamma \equiv kx, \quad \varphi(x) = \frac{\sum_\gamma n_\gamma v_{\gamma\gamma} f_\gamma(x)}{\sum_\lambda n_\lambda v_{\lambda\lambda} \Theta_\lambda^{-1} f_\lambda(x)}. \tag{A10}$$

Let us separate $I_\beta$ into two terms: $I_\beta = I_\beta^+ - I_\beta^-$, where

$$I_\beta^\pm = \frac{1}{2\pi i k} \int_{-\infty}^{+\infty} \frac{dx}{x - x_\beta - i\Delta} \cdot \frac{\varphi(x)}{x} \left[ \frac{1}{\varepsilon^\pm(kx,\mathbf{k})} - \frac{1}{\varepsilon(0,\mathbf{k})} \right], \tag{A11}$$

$$\varepsilon(0,\mathbf{k}) = 1 + (kL)^{-2}.$$

Since functions $\varepsilon^+$ and $\varepsilon^-$ are analytical in the upper and lower half-planes, respectively, the integrals can be calculated by the theory of residues. Further, we will consider two-component plasma with $n_i = n_e \equiv n$. After cumbersome calculations, we obtain

$$I_\beta = \frac{\varphi(x_\beta)}{kx_\beta} \left[ \frac{1}{\varepsilon^+(kx,\mathbf{k})} - \frac{1}{\varepsilon(0,\mathbf{k})} \right] + (\Theta_e - \Theta_i) S_\beta. \tag{A12}$$

The second term arose due to the poles of the function $\varphi(x)$ at $\Theta_e \neq \Theta_i$ (the precise expression for $S_\beta$ is not given because of its complexity). The Debye formula for

$$g_{\alpha\beta}(\mathbf{k}) \equiv \int d\mathbf{v}_\alpha d\mathbf{v}_\beta g_{\alpha\beta}(\mathbf{k},\mathbf{v}_\alpha,\mathbf{v}_\beta)$$

is easily obtained from eqs (A4), (A5), (A10), and (A12).

At $\Theta_e \gg \Theta_i$, we obtain the following expressions for the correlation functions:

$$k_i L_i = (2|\ln \mu|)^{-1/2}, \quad \text{sign}\, x = \frac{x}{|x|}, \tag{A13}$$

$$n g_{ii}(k) = -\left[1 + (kL_i)^2\right]^{-1} + \frac{1}{2}\left[1 - \text{sign}(k - k_i)\right]\left[1 + (kL_e)^2\right], \tag{A14}$$

$$n g_{ee}(k) = -\frac{\Theta_i}{2\Theta_e}\left[1 + (kL_i)^2\right]^{-1} - \frac{1}{2}\text{sign}(k - k_i)\left[1 + (kL_e)^2\right]^{-1}, \tag{A15}$$

$$n g_{ie}(k) = \frac{\Theta_i}{\Theta_e}\left[1 + (kL_i)^2\right]^{-1} + \frac{1}{2}\left[1 - \text{sign}(k - k_i)\right]. \tag{A16}$$

The additional term in energy due to the Coulomb interaction is expressed in terms of the correlation functions as

$$E_\text{coul} = \frac{1}{2} \int \frac{d\mathbf{k}}{(2\pi)^3} \sum_{\alpha,\beta} n_\alpha n_\beta v_{\alpha\beta}(\mathbf{k}) g_{\alpha\beta}(\mathbf{k}). \tag{A17}$$

Inserting (A4)-(A6), we obtain

$$E_\text{coul} = \frac{1}{2} \int \frac{d\mathbf{k}}{(2\pi)^3} \sum_\alpha n_\alpha v_{\alpha\alpha}(\mathbf{k}) \int d\mathbf{v}_\alpha f_\alpha(\mathbf{v}_\alpha) \frac{1 - \varepsilon^+(\mathbf{k}\mathbf{v}_\alpha,\mathbf{k})\varepsilon^-(\mathbf{k}\mathbf{v}_\alpha,\mathbf{k})}{\varepsilon^+(\mathbf{k}\mathbf{v}_\alpha,\mathbf{k})\varepsilon^-(\mathbf{k}\mathbf{v}_\alpha,\mathbf{k})}. \tag{A18}$$

Using identity (A8), we can again rewrite this equation in the form suitable for treating the limiting cases; the method of calculations is similar to the above-described one. Using eq (15), we obtain the final result:

$$\Delta P = -\frac{1}{24\pi L^3}\left[\frac{\Theta_e + \mu\Theta_i}{1+\mu} + 2(\Theta_e - \Theta_i)\sum_{n=0}^{\infty}\mathrm{Re}\frac{\Phi_n^{3/2}}{\ln\mu + i\pi(2n+1)}\right],$$

$$\Phi_n = \sum_{\alpha}\left(\frac{L}{L_\alpha}\right)^2 \int_{-\infty}^{+\infty}\frac{xf_\alpha(x)dx}{x - a_n - ib_n}, \qquad (A19)$$

$$\left.\begin{array}{l}a_n\\b_n\end{array}\right\} = \sqrt{\pm\frac{\ln\mu}{u_i^{-2} - u_e^{-2}} + \frac{1}{|u_i^{-2} - u_e^{-2}|}\sqrt{\ln^2\mu + \pi^2(2n+1)^2}}.$$

From two values of the root, $\sqrt{\Phi_n}$, the one is taken that has the positive imaginary part.

When $\Theta_e \gg \Theta_i$, the major contribution to pressure comes from $g_{ii}(k)$. From eq (A19), the following result is obtained:

$$\Delta P = \frac{1}{36\pi^2}\Theta_e k_i^3. \qquad (A20)$$

# CORRELATION FUNCTIONS AND PRESSURE OF A NONISOTHERMAL PLASMA

E. S. Medvedev

Expressions are obtained for the correlation functions of a classical nonisothermal two-component plasma. In the limiting case $\Theta_e \gg \Theta_i$ strong correlations arise because of the presence of weakly-damped waves (ion acoustic waves) with a phase velocity lying between the thermal velocities of the particles $u_e \gg \omega/k \gg u_i$. Under such conditions the correction to the pressure because of Coulomb interaction changes sign in comparison with the case of thermodynamic equilibrium, when there are no such waves.

The present paper describes a study of the effect of weakly-damped waves on the correlation functions of a nonisothermal plasma. Although the method employed (the test particle method [1]), is not a rigorous one, it is both simple and readily understandable physically. In a step-by-step attack on the problem it is necessary to solve an equation for the binary correlation function. In the classical case an equation of this type relating to correlations in nonequilibrium systems was obtained by Bogolyubov [2]. It was later extended to the quantum case by Klimontovich and Temko [3], and was subsequently solved by Silin [4]. The solution of the Bogolyubov equation is analyzed in the Appendix to the present paper.

The potential of the electric field excited in a plasma by a moving charge of type $\beta$ satisfies the Poisson equation

$$\Delta \varphi(r) = -4\pi e_\beta \delta(r - x(t)) - 4\pi \sum_\alpha e_\alpha \delta n_{\alpha\beta}(r, x(t)). \quad (1)$$

Here $\delta n_{\alpha\beta}(r, x(t))$ is the perturbation in the charge density of charges of type $\alpha$ at the point r. The set of hydrodynamic equations for particles of type $\alpha$ has the following form:

$$m_\alpha n_\alpha \frac{\partial v_\alpha}{\partial t} = -\nabla p_\alpha - e_\alpha n_\alpha \nabla \varphi, \quad (2a)$$

$$\frac{\partial n_\alpha}{\partial t} + n_\alpha \nabla v_\alpha = 0, \quad (2b)$$

$$\frac{d}{dt}\left(\frac{p_\alpha}{n_\alpha^a}\right) = 0; \quad p_\alpha = n_{\alpha 0}\Theta_\alpha. \quad (2c)$$

Here $n_{\alpha 0}$ and $\Theta_\alpha$ denote the mean density and the temperature of the particles of type $\alpha$, and $a$ is a constant of the order of unity. On linearizing set (2) and introducing the Fourier transform of (1), we obtain the following set of equations for the quantities $\delta n_{\alpha\beta}(k, x(t))$:

$$\frac{\partial^2}{\partial t^2}\delta n_{\alpha\beta} + a(\kappa u_\alpha)^2 \delta n_{\alpha\beta} +$$

$$+ \omega_\alpha^2 (\sum_\gamma z_{\alpha\gamma}\delta n_{\gamma\beta} + z_{\alpha\beta}e^{-ikx(t)}) = 0,$$

$$\omega_\alpha^2 = \frac{4\pi n_{\alpha 0}e_\alpha^2}{m_\alpha}; \quad u_\alpha^2 = \frac{\Theta_\alpha}{m_\alpha}; \quad z_{\alpha\beta} = \frac{e_\beta}{e_\alpha}. \quad (3)$$

We assume that the plasma particles have no inverse effect on the motion of of the test particle, so that $x(t) = x_0 + vt$. The solution of (3) is then given by the formula

$$\delta n_{\alpha\beta}(k, x(t)) = \frac{z_{\alpha\beta}}{\epsilon(kv, k)} \frac{\omega_\alpha^2}{(kv)^2 - a(\kappa u_\alpha)^2} e^{-ikx(t)}, \quad (4)$$

$$\epsilon(\omega, k) = 1 - \sum_\gamma \frac{\omega_\gamma^2}{\omega^2 - a(\kappa u_\gamma)^2}. \quad (5)$$

Let us now work out the density correlation function. By definition

$$n_{\alpha 0}n_{\gamma 0}g_{\alpha\gamma}(r, r') =$$
$$= \sum_\beta n_{\beta 0} \int dx\, dv f_\beta(v)[n_\alpha(r) - n_{\alpha 0}][n_\gamma(r') - n_{\gamma 0}], \quad (6)$$

where $f_\beta(v)$ denotes the velocity distribution function normalized to unity, and

$$n_\alpha(r) = n_{\alpha 0} + \delta_{\alpha\beta}\delta(r - x(t)) + \delta n_{\alpha\beta}(r, x(t)). \quad (7)$$

It can be shown by substituting (7) into (6), utilizing (4) and (5) and integrating with respect to x, that $g_{\alpha\gamma}(r, r')$ depends only on $r - r'$. We obtain for the Fourier components

$$n_\alpha n_\gamma g_{\alpha\gamma}(k) =$$
$$= \sum_\beta n_\beta \int_{-\infty}^{+\infty} du\, f_\beta(u) \left[\delta_{\alpha\beta} + \frac{z_{\alpha\beta}}{\kappa^2 \epsilon(\kappa u, k)} \frac{\omega_\alpha^2}{u^2 - au_\alpha^2}\right] \times$$
$$\times \left[\delta_{\gamma\beta} + \frac{z_{\gamma\beta}}{\kappa^2 \epsilon(-\kappa u, -k)} \frac{\omega_\gamma^2}{u^2 - au_\gamma^2}\right] \quad (8)$$

(here the integration with respect to velocity components perpendicular to k has been carried out, $f_\beta(u) = \int dv f_\beta(v)\delta\left(u - \frac{\kappa v}{\kappa}\right)$; the index 0 has been dropped from the notation). Since the final results depend only weakly on the constant $a$, we shall in future put $a = 1$. The integral in (8) diverges at points where $\epsilon(ku, k)$ reduces to zero. A discussion along the same lines as above but based upon the kinetic equation shows that this divergence is removed when allowance is made for the damping of the oscillations. After having added the correct imaginary part to $\epsilon(ku, k)$ [1], we investigate the effect of plasma oscillations on the correlation functions in the limiting case $\Theta_e \gg \Theta_i$.

On solving the equation $\epsilon(\omega, k) = 0$, we find that there are two branches of oscillations: a Langmuir branch with a frequency $\omega^2 = \omega_e^2 + k^2 u_e^2$, and an ion-acoustic branch with $\omega^2 = \omega_i^2 [1 + (kL_e)^{-2}]^{-1}$ ($L_e =$ $(\Theta_e/4\pi n_e e^2)]$. The Langmuir oscillations are weakly damped when $kL_e \ll 1$, but for such values of k their phase velocity is very much greater than the electron thermal velocity $u_e$. The number of particles participating in the excitation of this branch is consequently exponentially small, so that the associated contribution to (8) may be neglected. The situation is different with the ion-acoustic oscillations. The phase velocity in this case satisfies the inequalities $u_e \gg \omega/k \gg u_i$. This means on the one hand that the damping is small, and on the other that this oscillation branch is excited by a large group of particles (all electrons). It will be shown below that, in comparison with the case of statistical equilibrium, a considerable change takes place in the correlations.

For simplicity let us suppose that the plasma consists of electrons and of ions of a single type, and also that $n_i = n_e \equiv n$. In accordance with the above, the main contribution to (8) comes from the region $u_e > u > u_i$. In this interval we have

$$\epsilon(\kappa u, \mathbf{k}) \equiv \epsilon^*(-\kappa u, -\mathbf{k}) = 1 + (\kappa L_e)^{-2} - \left(\frac{\omega_i}{\kappa u}\right)^2 - i\Gamma(u), \quad (9)$$

where $\Gamma(u) = \sqrt{\pi}\frac{u}{u_e}(\kappa L_e)^{-2}$. Since $\Gamma^2 \ll |\epsilon|^2$ everywhere with the exception of the point $u = u_0$ corresponding to ion sound, we put $\Gamma(u) \equiv \Gamma(u_0)$. Suppose first of all that $\alpha = \gamma = i$. We then obtain from (8) and (9)

$$ng_{ii}(\kappa) = \frac{1}{u_e}\sqrt{\frac{2}{\pi}} \int_{u_i}^{u_e} du \left(\frac{\omega_i}{\kappa u}\right)^4 \left\{\left[1 + (\kappa L_e)^{-2} - \left(\frac{\omega_i}{\kappa u}\right)^2\right]^2 + \pi(\kappa L_e)^{-4}\left(\frac{u_0}{u_e}\right)^2\right\}^{-1}. \quad (10)$$

Since the integral converges, integration may be extended over the entire right semiaxis. After a few straightforward manipulations, relationship (10) takes the form

$$ng_{ii}(\kappa) = \frac{1}{u_e}\sqrt{\frac{2}{\pi}} \int_0^\infty \left\{\left[\left(\frac{u}{u_0}\right)^2 - 1\right]^2 + \frac{\pi}{\mu^2}\left(\frac{u_0}{u_i}\right)^6\right\}^{-1} du; \quad \mu = \frac{u_e \Theta_e}{u_i \Theta_i}. \quad (11)$$

Finally

$$ng_{ii}(\kappa) = 1 + (\kappa L_e)^2, \quad \kappa L_i \ll 1. \quad (12)$$

When $kL_i \gg 1$ the quantity $u_0$ becomes less than $u_i$; the ion sound begins to experience heavy damping and its contribution to $g_{ii}(k)$ falls off as $(kL_i)^{-4}$. The following expressions are obtained in like manner for the two other correlation functions:

$$ng_{ee}(\kappa) = \frac{1}{2}[1 + (\kappa L_e)^2]^{-1}, \quad ng_{ie}(\kappa) = 1 \quad (\kappa L_i \ll 1). \quad (13)$$

We recall that the Debye formula

$$g_{\alpha\beta}(\kappa) = -\frac{v_{\alpha\beta}(\kappa)}{\Theta}[1 + (\kappa L)^{-2}]^{-1},$$

$$v_{\alpha\beta}(\kappa) = \frac{4\pi e_\alpha e_\beta}{\kappa^2}, \quad L^{-2} = \frac{\sum_\gamma 4\pi n_\gamma e_\gamma^2}{\Theta} \quad (14)$$

is valid for the case of thermal equilibrium. We mention that similar terms are present in the precise expressions [see Appendix, formulas (A14)–(A16)], although they are small when $kL_i \ll 1$ (more precisely when $k \leq k_i$, $k_i L_i = (2|\ln\mu|)^{-1/2}$

A comparison of (14) with (13) shows that correlations originating from waves, unlike Debye correlations, do not at all follow the rule "like particles repel, unlike particles attract;" it is possible for like particles to attract. This can be understood when it is remembered that the density fluctuations at two nearby points A and B induced by a remote electron must be the same because of the homogeneity of the system; if, for example, the electrons at point A are compressed, the electron density at B also increases, which corresponds to an attraction (under conditions of thermal equilibrium this effect is swamped by direct Coulomb repulsion). The type of wave responsible for the perturbation is immaterial; a similar conclusion is reached from a study of the correlations produced by unlike particles depends, however, on the nature of the waves. In an ion acoustic wave div j = 0. This means that the electron and ion currents are oppositely directed and, consequently, that the particles themselves are moving in the one direction. Consequently, compression of the electrons at a given point is accompanied by the compression of the ions, i.e., there is attraction. In Langmuir oscillations, on the other hand, the currents are in the same direction, which leads to a change in the sign.

Formulas (12) and (13) have the further interesting feature that in absolute magnitude $g_{ii}(k)$ is large compared with $g_{ie}(k)$ and $g_{ee}(k)$ in the most important region $L_e^{-1} \ll k \ll L_i^{-1}$. The electrons exciting the ion sound move with velocities much in excess of the wave phase velocity. The emitted wave has the form of a cone with an apex angle of the order of $\omega/ku_e \ll 1$. The electron is thus correlated only with those particles falling within this cone. The correlations of the ions among themselves are not subject to such a restriction.

Let us find the correction to the pressure necessitated by Coulomb interaction. From the virial theorem

$$\Delta P \equiv P - n(\Theta_e + \Theta_i) = \frac{1}{3}E_{\text{coul}}. \quad (15)$$

In formula (A17) for $E_{\text{coul}}$ only the one term with $g_{ii}(k)$ need be retained. Integrating from k to $k_i$, we obtain

$$\Delta P = \frac{1}{36\pi^2}\Theta_e \kappa_i^3, \quad (16)$$

which agrees with (A20). It can be seen that the effective interaction between the particles which arises because of the presence of the undamped slow waves changes the sign of $\Delta P$ in comparison with the case of thermodynamic equilibrium.

The author wishes to thank A. A. Vedenov for suggesting the problem and for much valuable assistance.

APPENDIX

The Bogolyubov equation for the first term in the expansion of the binary correlation function in powers of $N_D^{-1}$ ($N_D$ is the number of particles in the Debye sphere) has the form ([2], see also [5])

$$n_\alpha n_\beta \nu_{\alpha\beta}(k) g_{\alpha\beta}(k, v_\alpha, v_\beta) = (kv_\alpha - kv_\beta - i\Delta)^{-1} \times$$

$$\times \left[ n_\alpha n_\beta f_\alpha(v_\alpha) f_\beta(v_\beta) \times \nu_{\alpha\beta}^2(k) \left( \frac{kv_\beta}{\Theta_\beta} - \frac{kv_\alpha}{\Theta_\alpha} \right) - \right.$$

$$- n_\alpha \nu_{\alpha\alpha}(k) f_\alpha(v_\alpha) \frac{kv_\alpha}{\Theta_\alpha} h_\beta(-k, v_\beta) +$$

$$\left. + n_\beta \nu_{\beta\beta}(k) f_\beta(v_\beta) \frac{kv_\beta}{\Theta_\beta} h_\alpha(k, v_\alpha) \right], \quad \Delta \to 0, \quad (A1)$$

$$g_{\alpha\beta}(r, v_\alpha, v_\beta) = \int \frac{dk}{(2\pi)^3} e^{ikr} g_{\alpha\beta}(k, v_\alpha, v_\beta), \quad (A2)$$

$$g_{\alpha\beta}(r, v_\alpha, v_\beta) \equiv F_{\alpha\beta}(r, v_\alpha, v_\beta) - 1,$$

$$h_\alpha(k, v_\alpha) = \sum_\beta n_\alpha n_\beta \nu_{\alpha\beta}(k) \int dv_\beta g_{\alpha\beta}(k, v_\alpha, v_\beta),$$

$$\nu_{\alpha\beta}(k) = \frac{4\pi e_\alpha e_\beta}{k^2},$$

$$f_\alpha(v_\alpha) = \left( \frac{m_\alpha}{2\pi\Theta_\alpha} \right)^{3/2} \exp\left( -\frac{m_\alpha v_\alpha^2}{2\Theta_\alpha} \right). \quad (A3)$$

Here $F_{\alpha\beta}(r, v_\alpha, v_\beta)$ denotes the binary correlation function. The method presented in [4] for the solution of the corresponding quantum equation is directly applicable to the present case. On carrying out the appropriate calculations, we obtain

$$g_{\alpha\beta}(k, v_\alpha, v_\beta) = q_{\alpha\beta}(k, v_\alpha, v_\beta) + q_{\beta\alpha}^*(k, v_\beta, v_\alpha), \quad (A4)$$

$$q_{\alpha\beta}(k, v_\alpha, v_\beta) =$$

$$= -\frac{\nu_{\alpha\beta} f_\alpha(v_\alpha) f_\beta(v_\beta)}{kv_\alpha - kv_\beta - i\Delta} \cdot \frac{kv_\alpha}{\Theta_\alpha} \left[ \frac{1}{\epsilon^+(kv_\beta, k)} - \frac{kv_\beta}{\Theta_\beta} I_\beta \right], \quad (A5)$$

$$I_\beta = \sum_\gamma n_\gamma \nu_{\gamma\gamma} \int \frac{dv_\gamma f_\gamma(v_\gamma)}{kv_\beta - kv_\gamma + i\Delta} \cdot$$

$$\cdot [\epsilon^+(kv_\gamma, k) \epsilon^-(kv_\gamma, k)]^{-1}, \quad (A6)$$

$$\epsilon^\pm(\omega, k) \equiv \epsilon(\omega \pm i\Delta, k);$$

$$\epsilon(\omega, k) = 1 + \sum_\alpha n_\alpha \nu_{\alpha\alpha} \int \frac{dv_\alpha}{\omega - kv_\alpha} \left[ k \frac{\partial f_\alpha(v_\alpha)}{m_\alpha \partial v_\alpha} \right]. \quad (A7)$$

We change the form of $I_\beta$ by means of the following identity:

$$\frac{1}{\epsilon^+(\omega, k) \epsilon^-(\omega, k)} \equiv$$

$$\equiv \frac{1}{2 \mathrm{Im}\, \epsilon^-(\omega, k)} \left[ \frac{1}{\epsilon^+(\omega, k)} - \frac{1}{\epsilon^-(\omega, k)} \right]. \quad (A8)$$

Then (A6) takes the form

$$I_\beta = \frac{1}{2\pi i \kappa} \int_{-\infty}^{+\infty} \frac{dx}{x - x_\beta - i\Delta} \cdot$$

$$\cdot \frac{\varphi(x)}{x} \left[ \frac{1}{\epsilon^+(\kappa x, k)} - \frac{1}{\epsilon^-(\kappa x, k)} \right], \quad (A9)$$

$$kv_\beta \equiv \kappa x_\beta, \quad kv_\gamma \equiv \kappa x, \quad \varphi(x) = \frac{\sum_\gamma n_\gamma \nu_{\gamma\gamma} f_\gamma(x)}{\sum_\lambda n_\lambda \nu_{\lambda\lambda} \Theta_\lambda^{-1} f_\lambda(x)}. \quad (A10)$$

We decompose $I_\beta$ into two components $I_\beta = I_\beta^+ - I_\beta^-$;

$$I_\beta^\pm = \frac{1}{2\pi i \kappa} \int_{-\infty}^{+\infty} \frac{dx}{x - x_\beta - i\Delta} \cdot$$

$$\cdot \frac{\varphi(x)}{x} \left[ \frac{1}{\epsilon^\pm(\kappa x, k)} - \frac{1}{\epsilon(0, k)} \right],$$

$$\epsilon(0, k) = 1 + (\kappa L)^{-2}. \quad (A11)$$

Since the functions $\epsilon^+$ and $\epsilon^-$ are analytic in the upper and lower half planes respectively, the integrals may be worked out using the theory of residues. From now on we shall be considering a two-component plasma with $n_i = n_e \equiv n$. We obtain after some cumbersome calculations

$$I_\beta = \frac{\varphi(x_\beta)}{\kappa x_\beta} \left[ \frac{1}{\epsilon^+(\kappa x_\beta, k)} - \frac{1}{\epsilon(0, k)} \right] + (\Theta_e - \Theta_i) S_\beta. \quad (A12)$$

The second component comes about because of poles in the function $\varphi(x)$ at $\Theta_e \neq \Theta_i$ (the precise expression for $S_\beta$ is very cumbersome and will not be reproduced here). The Debye result for $g_{\alpha\beta}(k) \equiv \int dv_\alpha dv_\beta g_{\alpha\beta}(k, v_\alpha, v_\beta)$ is readily obtained from expressions (A4), (A5), (A10), and (A12).

When $\Theta_e \gg \Theta_i$, we obtain for the correlation functions

$$\kappa_i L_i = (2|\ln\mu|)^{-1/2}, \quad \mathrm{sign}\, x = \frac{x}{|x|}, \quad (A13)$$

$$n g_{ii}(\kappa) = -[1 + (\kappa L_i)^2]^{-1} +$$

$$+ \frac{1}{2} [1 - \mathrm{sign}(\kappa - \kappa_i)] [1 + (\kappa L_e)^2], \quad (A14)$$

$$n g_{ee}(\kappa) = -\frac{\Theta_i}{2\Theta_e} [1 + (\kappa L_i)^2]^{-1} -$$

$$- \frac{1}{2} \mathrm{sign}(\kappa - \kappa_i) [1 + (\kappa L_e)^2]^{-1}, \quad (A15)$$

$$n g_{ie}(\kappa) = \frac{\Theta_i}{\Theta_e} [1 + (\kappa L_i)^2]^{-1} + \frac{1}{2} [1 - \mathrm{sign}(\kappa - \kappa_i)]. \quad (A16)$$

The contribution to the energy of the system coming from Coulomb interaction is expressed in terms of the correlation functions in the following manner:

$$E_{\text{coul}} = \frac{1}{2} \int \frac{d\mathbf{k}}{(2\pi)^3} \sum_{\alpha,\beta} n_\alpha n_\beta \nu_{\alpha\beta}(\mathbf{k}) g_{\alpha\beta}(\mathbf{k}). \quad (A17)$$

Substituting (A4)–(A6) into this expression, we obtain

$$E_{\text{coul}} = \frac{1}{2} \int \frac{d\mathbf{k}}{(2\pi)^3} \sum_{\alpha} n_\alpha \nu_{\alpha\alpha}(\mathbf{k}) \times$$
$$\times \int d\mathbf{v}_\alpha f_\alpha(\mathbf{v}_\alpha) \frac{1 - \varepsilon^+(\mathbf{k}\mathbf{v}_\alpha, \mathbf{k})\varepsilon^-(\mathbf{k}\mathbf{v}_\alpha, \mathbf{k})}{\varepsilon^+(\mathbf{k}\mathbf{v}_\alpha, \mathbf{k})\varepsilon^-(\mathbf{k}\mathbf{v}_\alpha, \mathbf{k})}. \quad (A18)$$

This formula may be reduced using identity (A8) to a form more convenient for the study of limiting cases (the procedure is the same as before). The final result for the pressure correction is as follows [here we have utilized formula (15)]:

$$\Delta P = -\frac{1}{24\pi L^3} \left[ \frac{\Theta_e + \mu \Theta_i}{1 + \mu} + \right.$$
$$\left. + 2(\Theta_e - \Theta_i) \sum_{n=0}^{\infty} \text{Re} \frac{\Phi_n^{3/2}}{\ln \mu + i\pi(2n+1)} \right],$$

$$\Phi_n = \sum_\alpha \left(\frac{L}{L_\alpha}\right)^2 \int_{-\infty}^{+\infty} \frac{x f_\alpha(x) \, dx}{x - a_n - ib_n},$$

$$\left.\begin{array}{c} a_n \\ b_n \end{array}\right\} = \left[ \pm \frac{\ln \mu}{u_i^{-2} - u_e^{-2}} + \right.$$
$$\left. + \frac{1}{|u_i^{-2} - u_e^{-2}|} \sqrt{\ln^2 \mu + \pi^2 (2n+1)^2} \right]^{-1}. \quad (A19)$$

Of the two values of the square root $\sqrt{\Phi_n}$, that one is taken which has a positive imaginary part.

When $\Theta_e \gg \Theta_i$, the main contribution to the pressure comes from $g_{ii}(k)$. The following result is obtained from (A19):

$$\Delta P = (1/36\pi^2)(\Theta_e k_i^3). \quad (A20)$$

REFERENCES

1. M. N. Rosenbluth and N. Rostoker, Phys. Fluids, 5, 776, 1962.
2. N. N. Bogolyubov, Problems of Dynamical Theory in Statistical Physics [in Russian], Gostekhizdat 1946.
3. Yu. L. Klimontovich and S. V. Temko, ZhETF, 33, 132, 1957; Nauchn. dokl. vyssh. shkol., fiz.-mat. nauki, C. V. Temko, 2, 189, 1958.
4. V. P. Silin, ZhETF, 40, 1768, 1961.
5. A. Lenard, Ann. of Phys., 3, 390, 1960.

13 June 1966          Moscow Institute of Technical Physics

Э. С. МЕДВЕДЕВ

## КОРРЕЛЯЦИОННЫЕ ФУНКЦИИ И ДАВЛЕНИЕ НЕИЗОТЕРМИЧЕСКОЙ ПЛАЗМЫ

Получены выражения для корреляционных функций классической неизотермической двухкомпонентной плазмы. В предельном случае $\Theta_e \gg \Theta_i$ возникают сильные корреляции благодаря наличию слабозатухающих волн (ионного звука), фазовая скорость которых лежит между тепловыми скоростями частиц $u_e \gg \dfrac{\omega}{\kappa} \gg u_i$. При этом добавка к давлению за счет кулоновского взаимодействия меняет знак по сравнению со случаем термодинамического равновесия, когда такие волны отсутствуют.

В настоящей работе исследуется влияние слабозатухающих волн на корреляционные функции неизотермической плазмы. Используемый метод (метод пробной частицы [1]) не является строгим, однако он прост и физически нагляден. Последовательный подход требует изучения решения уравнения для бинарной корреляционной функции. В классическом случае такое уравнение, пригодное для исследования корреляций в неравновесных системах, было получено Боголюбовым [2]. Обобщенное на случай квантовой системы Климонтовичем и Темко [3], оно было решено Силиным [4]. В приложении дан анализ решения уравнения Боголюбова.

Потенциал электрического поля, возбуждаемого в плазме движущимся зарядом сорта β, удовлетворяет уравнению Пуассона:

$$\Delta\varphi(\boldsymbol{r}) = -4\pi e_\beta \delta(\boldsymbol{r} - \boldsymbol{x}(t)) - 4\pi \sum_\alpha e_\alpha \delta n_{\alpha\beta}(\boldsymbol{r}, \boldsymbol{x}(t)). \quad (1)$$

Здесь $\delta n_{\alpha\beta}(\boldsymbol{r}, \boldsymbol{x}(t))$ — возмущение плотности зарядов сорта α в точке $\boldsymbol{r}$. Система уравнений гидродинамики для частиц сорта α имеет следующий вид

$$m_\alpha n_\alpha \frac{\partial \boldsymbol{v}_\alpha}{\partial t} = -\nabla p_\alpha - e_\alpha n_\alpha \nabla \varphi, \quad (2а)$$

$$\frac{\partial n_\alpha}{\partial t} + n_\alpha \nabla \boldsymbol{v}_\alpha = 0, \quad (2б)$$

$$\frac{d}{dt}\left(\frac{p_\alpha}{n_\alpha^a}\right) = 0; \quad p_\alpha = n_{\alpha 0}\Theta_\alpha. \quad (2в)$$

Здесь $n_{\alpha 0}$ и $\Theta_\alpha$ — средняя плотность и температура частиц сорта α, $a$ — постоянная порядка единицы. Линеаризуя систему (2) и используя Фурье-образ (1), получим следующую систему уравнений для величин $\delta n_{\alpha\beta}(\boldsymbol{k}, \boldsymbol{x}(t))$:

$$\frac{\partial^2}{\partial t^2}\delta n_{\alpha\beta} + a(\kappa u_\alpha)^2 \delta n_{\alpha\beta} + \omega_\alpha^2\left(\sum_\gamma z_{\alpha\gamma}\delta n_{\gamma\beta} + z_{\alpha\beta} e^{-i\boldsymbol{k}\boldsymbol{x}(t)}\right) = 0,$$

$$(3)$$

$$\omega_\alpha^2 = \frac{4\pi n_{\alpha 0} e_\alpha^2}{m_\alpha}; \quad u_\alpha^2 = \frac{\Theta_\alpha}{m_\alpha}; \quad z_{\alpha\beta} = \frac{e_\beta}{e_\alpha}. \tag{3}$$

Пренебрегая обратным влиянием частиц плазмы на движение пробной частицы, положим $\boldsymbol{x}(t) = \boldsymbol{x}_0 + \boldsymbol{vt}$. Тогда решение (3) дается формулой

$$\delta n_{\alpha\beta}(\boldsymbol{k}, \boldsymbol{x}(t)) = \frac{z_{\alpha\beta}}{\in(\boldsymbol{kv}, \boldsymbol{k})} \frac{\omega_\alpha^2}{(\boldsymbol{kv})^2 - a(\varkappa u_\alpha)^2} e^{-i\boldsymbol{k}\boldsymbol{x}(t)}, \tag{4}$$

$$\in(\omega, \boldsymbol{k}) = 1 - \sum_\gamma \frac{\omega_\gamma^2}{\omega^2 - a(\varkappa u_\gamma)^2}. \tag{5}$$

Вычислим теперь корреляционную функцию плотности. По определению

$$n_{\alpha 0} n_{\gamma 0} g_{\alpha\gamma}(\boldsymbol{r}, \boldsymbol{r}') = \sum_\beta n_{\beta 0} \int d\boldsymbol{x}\, d\boldsymbol{v} f_\beta(\boldsymbol{v}) [n_\alpha(\boldsymbol{r}) - n_{\alpha 0}] [n_\gamma(\boldsymbol{r}') - n_{\gamma 0}], \tag{6}$$

где $f_\beta(\boldsymbol{v})$ — функция распределения, нормированная на единицу;

$$n_\alpha(\boldsymbol{r}) = n_{\alpha 0} + \delta_{\alpha\beta}\delta(\boldsymbol{r} - \boldsymbol{x}(t)) + \delta n_{\alpha\beta}(\boldsymbol{r}, \boldsymbol{x}(t)). \tag{7}$$

Подставляя (7) в (6) и используя (4) и (5), после интегрирования по $\boldsymbol{x}$ найдем, что $g_{\alpha\gamma}(\boldsymbol{r}, \boldsymbol{r}')$ зависит только от $\boldsymbol{r} - \boldsymbol{r}'$. Для компоненты Фурье получим

$$n_\alpha n_\gamma g_{\alpha\gamma}(\boldsymbol{k}) = \sum_\beta n_\beta \int_{-\infty}^{+\infty} du\, f_\beta(u) \left[\delta_{\alpha\beta} + \frac{z_{\alpha\beta}}{\varkappa^2 \in(\varkappa u, \boldsymbol{k})} \frac{\omega_\alpha^2}{u^2 - au_\alpha^2}\right] \times$$
$$\times \left[\delta_{\gamma\beta} + \frac{z_{\gamma\beta}}{\varkappa^2 \in(-\varkappa u, -\boldsymbol{k})} \frac{\omega_\gamma^2}{u^2 - au_\gamma^2}\right] \tag{8}$$

(здесь выполнено интегрирование по компонентам скорости, перпендикулярным $\boldsymbol{\varkappa}$, $f_\beta(u) = \int d\boldsymbol{v} f_\beta(\boldsymbol{v}) \delta\left(u - \frac{\boldsymbol{\varkappa v}}{\varkappa}\right)$; индекс 0 у плотности будем опускать). Значение постоянной $a$ в дальнейшем несущественно, так как ответ зависит от нее слабо, и мы положим $a = 1$. Интеграл в (8) расходится в точках, где $\in(\varkappa u, \boldsymbol{k})$ обращается в нуль. Рассмотрение, аналогичное проведенному, но основанное на кинетическом уравнении, показывает, что учет затухания колебаний устраняет расходимость. Мы исследуем влияние колебаний плазмы на корреляционные функции в предельном случае $\Theta_e \gg \Theta_i$, добавляя в $\in(\varkappa u, \boldsymbol{k})$ правильную мнимую часть [1].

Решая уравнение $\in(\omega, \boldsymbol{k}) = 0$, найдем, что имеется две ветви колебаний: лэнгмюровские с частотой $\omega^2 = \omega_e^2 + \varkappa^2 u_e^2$ и ионно-звуковые $\omega^2 = \omega_i^2 [1 + (\varkappa L_e)^{-2}]^{-1}$ $\left(L_e^2 = \frac{\Theta_e}{4\pi n_e e^2}\right)$. Лэнгмюровские колебания слабо затухают при $\varkappa L_e \ll 1$, но при таких $\varkappa$ их фазовая скорость $\frac{\omega}{\varkappa}$ гораздо больше тепловой скорости электронов $u_e$. Поэтому в возбуждении этой ветви колебаний принимает участие экспоненциально малое число частиц, вследствие чего соответствующим вкладом в (8) можно пренебречь. Иное дело ионный звук. Его фазовая скорость удовлетворяет неравенствам: $u_e \gg \dfrac{\omega}{\varkappa} \gg u_i$. Это означает, с одной стороны, малость затухания, а с другой — наличие большой группы частиц (все электроны), возбуждающих данную ветвь колебаний. Как мы увидим ниже, это приводит к существенному изменению корреляций по сравнению со случаем статистического равновесия.

Пусть для простоты плазма состоит из электронов и одного сорта ионов, причем $n_i = n_e \equiv n$. В соответствии с вышесказанным, основной вклад в (8) дает область $u_e > u > u_i$. В этом интервале

$$\epsilon(\varkappa u, \boldsymbol{k}) \equiv \epsilon^*(-\varkappa u, -\boldsymbol{k}) = 1 + (\varkappa L_e)^{-2} - \left(\frac{\omega_i}{\varkappa u}\right)^2 - i\Gamma(u), \qquad (9)$$

где $\Gamma(u) = \sqrt{\dfrac{\pi}{2}} \dfrac{u}{u_e} (\varkappa L_e)^{-2}$. Поскольку $\Gamma^2 \ll |\epsilon|^2$ везде, за исключением точки $u = u_0$, отвечающей ионному звуку, мы положим $\Gamma(u) \equiv \Gamma(u_0)$. Пусть сначала $\alpha = \gamma = i$. Тогда из (8) и (9) получим:

$$ng_{ii}(\varkappa) = \frac{1}{u_e} \sqrt{\frac{2}{\pi}} \int\limits_{u_i}^{u_e} du \left(\frac{\omega_i}{\varkappa u}\right)^4 \Bigg\{ \bigg[ 1 + (\varkappa L_e)^{-2} -$$
$$- \left(\frac{\omega_i}{\varkappa u}\right)^2 \bigg]^2 + \pi (\varkappa L_e)^{-4} \left(\frac{u_0}{u_e}\right)^2 \Bigg\}^{-1}. \qquad (10)$$

Интегрирование можно распространить на всю правую полуось в силу сходимости интеграла. После несложных преобразований (10) принимает вид

$$ng_{ii}(\varkappa) = \frac{1}{u_e} \sqrt{\frac{2}{\pi}} \int\limits_0^\infty \Bigg\{ \bigg[ \left(\frac{u}{u_0}\right)^2 - 1 \bigg]^2 + \frac{\pi}{\mu^2} \left(\frac{u_0}{u_i}\right)^6 \Bigg\}^{-1} du; \quad \mu = \frac{u_e \Theta_e}{u_i \Theta_i}. \qquad (11)$$

Окончательно

$$ng_{ii}(\varkappa) = 1 + (\varkappa L_e)^2, \quad \varkappa L_i \ll 1. \qquad (12)$$

При $\varkappa L_i \gg 1$ $u_0$ становится меньше $u_i$; ионный звук начинает сильно затухать, и его вклад в $g_{ii}(\varkappa)$ уменьшается, как $(\varkappa L_i)^{-4}$. Для двух других корреляционных функций аналогичным путем можно получить следующие выражения:

$$ng_{ee}(\varkappa) = \frac{1}{2} [1 + (\varkappa L_e)^2]^{-1}, \quad ng_{ie}(\varkappa) = 1 \quad (\varkappa L_i \ll 1). \qquad (13)$$

Напомним, что в случае теплового равновесия справедлива дебаевская формула:

$$g_{\alpha\beta}(\varkappa) = -\frac{\nu_{\alpha\beta}(\varkappa)}{\Theta} [1 + (\varkappa L)^{-2}]^{-1}, \quad \nu_{\alpha\beta}(\varkappa) = \frac{4\pi e_\alpha e_\beta}{\varkappa^2}, \quad L^{-2} = \frac{\sum\limits_\gamma 4\pi n_\gamma e_\gamma^2}{\Theta}. \qquad (14)$$

В соответствующих точных выражениях (см. приложение (П14) — (П16)) аналогичные члены также присутствуют, но они малы при $\varkappa L_i \ll 1$ (точнее, при $\varkappa \leqslant \varkappa_i$, $\varkappa_i L_i = (2|\ln \mu|)^{-1/2}$).

Сравнивая (14) с (12) и (13), мы видим, что корреляции, обусловленные волнами, в отличие от дебаевских, отнюдь не удовлетворяют правилу: „одноименные частицы отталкиваются, разноименные притягиваются" — появляется притяжение одинаковых частиц. Это можно понять, если учесть, что флуктуации плотности в двух близких точках $A$ и $B$, вызванные удаленным электроном, должны быть одинаковы в силу однородности системы; если, например, в точке $A$ произошло сгущение электронов, то и в точке $B$ плотность электронов возрастает, а это означает притяжение [1]. При этом безразлично, какие именно волны ответственны за возмущение: рассматривая, к примеру, корреляции, обусловленные лэнгмюровскими колебаниями, мы пришли бы к аналогичному выводу. Знак же корреляционной функции для разноименных частиц существенно зависит от характера волн. В ионно-звуковой волне $\operatorname{div} \boldsymbol{j} = 0$. Это означает, что токи электронов и ионов направлены навстречу друг другу, и, следовательно, сами частицы движутся в одну сторону. Поэтому сгущение электронов в данной точке сопровождается сгущением ионов, то есть имеет место притяжение. В лэнгмюровских же колебаниях токи направлены в одну сторону, что приводит к изменению знака.

Другой интересной особенностью формул (12) и (13) является то, что $g_{ii}(\varkappa)$ по абсолютной величине велика по сравнению с $g_{ie}(\varkappa)$ и $g_{ee}(\varkappa)$ в наиболее существенной области $L_e^{-1} \ll \varkappa \ll L_i^{-1}$. Электроны, возбуждающие ионный звук, движутся со скоростями, значительно превышающими фазовую скорость волн. Фронт испускаемой волны имеет вид конуса с углом раствора порядка $\dfrac{\omega}{\varkappa u_e} \ll 1$. Поэтому электрон коррелирует только с теми частицами, которые попадают в этот конус. Корреляции же ионов между собой не подвержены подобным ограничениям.

Найдем добавку к давлению, обусловленную кулоновским взаимодействием. По теореме вириала

$$\Delta P \equiv P - n(\Theta_e + \Theta_i) = \frac{1}{3} E_{\text{кул}}. \tag{15}$$

В формуле (П17) для $E_{\text{кул}}$ можно оставить один только член с $g_{ii}(\varkappa)$. Интегрируя по $\varkappa$ до $\varkappa_i$, получим

$$\Delta P = \frac{1}{36\pi^2} \Theta_e \varkappa_i^3, \tag{16}$$

что совпадает с (П20). Мы видим, что эффективное взаимодействие частиц, возникающее благодаря наличию незатухающих медленных волн, приводит к изменению знака $\Delta P$ по сравнению со случаем термодинамического равновесия.

Автор благодарен А. А. Веденову за предложенную задачу и большую помощь.

Приложение

Уравнение Боголюбова для первого члена разложения бинарной корреляционной функции по $N_D^{-1}$ ($N_D$ — число частиц в дебаевской сфере) имеет вид ([2]; см. также [5]):

$$n_\alpha n_\beta \nu_{\alpha\beta}(\boldsymbol{k}) g_{\alpha\beta}(\boldsymbol{k}, \boldsymbol{v}_\alpha, \boldsymbol{v}_\beta) = (\boldsymbol{k}\boldsymbol{v}_\alpha - \boldsymbol{k}\boldsymbol{v}_\beta - i\Delta)^{-1} \bigg[ n_\alpha n_\beta f_\alpha(\boldsymbol{v}_\alpha) f_\beta(\boldsymbol{v}_\beta) \times$$

$$\times \nu_{\alpha\beta}^2(\boldsymbol{k}) \left( \frac{\boldsymbol{k}\boldsymbol{v}_\beta}{\Theta_\beta} - \frac{\boldsymbol{k}\boldsymbol{v}_\alpha}{\Theta_\alpha} \right) - n_\alpha \nu_{\alpha\alpha}(\boldsymbol{k}) f_\alpha(\boldsymbol{v}_\alpha) \frac{\boldsymbol{k}\boldsymbol{v}_\alpha}{\Theta_\alpha} h_\beta(-\boldsymbol{k}, \boldsymbol{v}_\beta) + \tag{П1}$$

---

[1] При тепловом равновесии прямое кулоновское отталкивание значительно превышает указанный эффект.

$$+ n_\beta \nu_{\beta\beta}(\boldsymbol{k}) f_\beta(\boldsymbol{v}_\beta) \frac{\boldsymbol{k v}_\beta}{\Theta_\beta} h_\alpha(\boldsymbol{k}, \boldsymbol{v}_\alpha)\Big], \quad \Delta \to 0,$$

$$g_{\alpha\beta}(\boldsymbol{r}, \boldsymbol{v}_\alpha, \boldsymbol{v}_\beta) = \int \frac{d\boldsymbol{k}}{(2\pi)^3} e^{i\boldsymbol{k r}} g_{\alpha\beta}(\boldsymbol{k}, \boldsymbol{v}_\alpha, \boldsymbol{v}_\beta), \quad (\text{П2})$$

$$g_{\alpha\beta}(\boldsymbol{r}, \boldsymbol{v}_\alpha, \boldsymbol{v}_\beta) \equiv F_{\alpha\beta}(\boldsymbol{r}, \boldsymbol{v}_\alpha, \boldsymbol{v}_\beta) - 1,$$

$$h_\alpha(\boldsymbol{k}, \boldsymbol{v}_\alpha) = \sum_\beta n_\alpha n_\beta \nu_{\alpha\beta}(\boldsymbol{k}) \int d\boldsymbol{v}_\beta g_{\alpha\beta}(\boldsymbol{k}, \boldsymbol{v}_\alpha, \boldsymbol{v}_\beta), \quad (\text{П3})$$

$$\nu_{\alpha\beta}(\boldsymbol{k}) = \frac{4\pi e_\alpha e_\beta}{k^2}, \quad f_\alpha(\boldsymbol{v}_\alpha) = \left(\frac{m_\alpha}{2\pi\Theta_\alpha}\right)^{3/2} \exp\left(-\frac{m_\alpha v_\alpha^2}{2\Theta_\alpha}\right).$$

Здесь $F_{\alpha\beta}(\boldsymbol{r}, \boldsymbol{v}_\alpha, \boldsymbol{v}_\beta)$ — бинарная корреляционная функция. Метод, использованный в [4] для решения соответствующего квантового уравнения, без всяких изменений пригоден и в данном случае, поэтому мы сразу выпишем ответ

$$g_{\alpha\beta}(\boldsymbol{k}, \boldsymbol{v}_\alpha, \boldsymbol{v}_\beta) = q_{\alpha\beta}(\boldsymbol{k}, \boldsymbol{v}_\alpha, \boldsymbol{v}_\beta) + q_{\beta\alpha}^*(\boldsymbol{k}, \boldsymbol{v}_\beta, \boldsymbol{v}_\alpha), \quad (\text{П4})$$

$$q_{\alpha\beta}(\boldsymbol{k}, \boldsymbol{v}_\alpha, \boldsymbol{v}_\beta) = -\frac{\nu_{\alpha\beta} f_\alpha(\boldsymbol{v}_\alpha) f_\beta(\boldsymbol{v}_\beta)}{\boldsymbol{k v}_\alpha - \boldsymbol{k v}_\beta - i\Delta} \cdot \frac{\boldsymbol{k v}_\alpha}{\Theta_\alpha} \left[\frac{1}{\varepsilon^+(\boldsymbol{k v}_\beta, \boldsymbol{k})} - \frac{\boldsymbol{k v}_\beta}{\Theta_\beta} I_\beta\right], \quad (\text{П5})$$

$$I_\beta = \sum_\gamma n_\gamma \nu_{\gamma\gamma} \int \frac{d\boldsymbol{v}_\gamma f_\gamma(\boldsymbol{v}_\gamma)}{\boldsymbol{k v}_\beta - \boldsymbol{k v}_\gamma + i\Delta} [\varepsilon^+(\boldsymbol{k v}_\gamma, \boldsymbol{k}) \varepsilon^-(\boldsymbol{k v}_\gamma, \boldsymbol{k})]^{-1}, \quad (\text{П6})$$

$$\varepsilon^\pm(\omega, \boldsymbol{k}) \equiv \varepsilon(\omega \pm i\Delta, \boldsymbol{k}); \quad \varepsilon(\omega, \boldsymbol{k}) = 1 + \sum_\alpha n_\alpha \nu_{\alpha\alpha} \int \frac{d\boldsymbol{v}_\alpha}{\omega - \boldsymbol{k v}_\alpha} \left[\boldsymbol{k} \frac{\partial f_\alpha(\boldsymbol{v}_\alpha)}{m_\alpha \partial \boldsymbol{v}_\alpha}\right]. \quad (\text{П7})$$

Преобразуем $I_\beta$ с помощью следующего тождества

$$\frac{1}{\varepsilon^+(\omega, \boldsymbol{k}) \varepsilon^-(\omega, \boldsymbol{k})} \equiv \frac{1}{2\operatorname{Im} \varepsilon^-(\omega, \boldsymbol{k})} \left[\frac{1}{\varepsilon^+(\omega, \boldsymbol{k})} - \frac{1}{\varepsilon^-(\omega, \boldsymbol{k})}\right]. \quad (\text{П8})$$

Тогда (П6) примет вид:

$$I_\beta = \frac{1}{2\pi i \kappa} \int_{-\infty}^{\infty} \frac{dx}{x - x_\beta - i\Delta} \cdot \frac{\varphi(x)}{x} \left[\frac{1}{\varepsilon^+(\kappa x, \boldsymbol{k})} - \frac{1}{\varepsilon^-(\kappa x, \boldsymbol{k})}\right]; \quad (\text{П9})$$

$$\boldsymbol{k v}_\beta \equiv \kappa x_\beta, \quad \boldsymbol{k v}_\gamma \equiv \kappa x, \quad \varphi(x) = \frac{\sum_\gamma n_\gamma \nu_{\gamma\gamma} f_\gamma(x)}{\sum_\lambda n_\lambda \nu_{\lambda\lambda} \Theta_\lambda^{-1} f_\lambda(x)}. \quad (\text{П10})$$

Разобьем $I_\beta$ на два слагаемых: $I_\beta = I_\beta^+ - I_\beta^-$ ;

$$I_\beta^\pm = \frac{1}{2\pi i \kappa} \int_{-\infty}^{+\infty} \frac{dx}{x - x_\beta - i\Delta} \cdot \frac{\varphi(x)}{x} \left[\frac{1}{\varepsilon^\pm(\kappa x, \boldsymbol{k})} - \frac{1}{\varepsilon(0, \boldsymbol{k})}\right], \quad (\text{П11})$$

$$\varepsilon(0, \boldsymbol{k}) = 1 + (\kappa L)^{-2}.$$

Так как функции $\varepsilon^+$ и $\varepsilon^-$ аналитичны соответственно в верхней и нижней полуплоскостях, интегралы можно вычислить с помощью теории вычетов. Далее будем рассматривать двухкомпонентную плазму с $n_i = n_e \equiv n$. После громоздких вычислений получаем

$$I_\beta = \frac{\varphi(x_\beta)}{\kappa x_\beta} \left[\frac{1}{\varepsilon^+(\kappa x_\beta, \boldsymbol{k})} - \frac{1}{\varepsilon(0, \boldsymbol{k})}\right] + (\Theta_e - \Theta_i) S_\beta. \quad (\text{П12})$$

Второе слагаемое возникло благодаря наличию полюсов у функции $\varphi(x)$ при $\Theta_e \neq \Theta_i$ (точный вид $S_\beta$ не приводим из-за его громоздкости). Из (П4), (П5), (П10) и (П12) легко получается дебаевский результат для $g_{\alpha\beta}(\mathbf{k}) \equiv \int d\mathbf{v}_\alpha d\mathbf{v}_\beta g_{\alpha\beta}(\mathbf{k}, \mathbf{v}_\alpha, \mathbf{v}_\beta)$.

При $\Theta_e \gg \Theta_i$ для корреляционных функций получим:

$$\kappa_i L_i = (2|\ln\mu|)^{-1/2}, \quad \text{sign}\, x = \frac{x}{|x|}, \tag{П13}$$

$$ng_{ii}(\kappa) = -[1+(\kappa L_i)^2]^{-1} + \frac{1}{2}[1-\text{sign}(\kappa-\kappa_i)][1+(\kappa L_e)^2], \tag{П14}$$

$$ng_{ee}(\kappa) = -\frac{\Theta_i}{2\Theta_e}[1+(\kappa L_i)^2]^{-1} - \frac{1}{2}\text{sign}(\kappa-\kappa_i)[1+(\kappa L_e)^2]^{-1}, \tag{П15}$$

$$ng_{ie}(\kappa) = \frac{\Theta_i}{\Theta_e}[1+(\kappa L_i)^2]^{-1} + \frac{1}{2}[1-\text{sign}(\kappa-\kappa_i)]. \tag{П16}$$

Добавка к энергии системы за счет кулоновского взаимодействия выражается через корреляционные функции следующим образом:

$$E_{\text{кул}} = \frac{1}{2}\int \frac{d\mathbf{k}}{(2\pi)^3} \sum_{\alpha,\beta} n_\alpha n_\beta \nu_{\alpha\beta}(\mathbf{k}) g_{\alpha\beta}(\mathbf{k}). \tag{П17}$$

Подставляя сюда (П4) — (П6), получим

$$E_{\text{кул}} = \frac{1}{2}\int \frac{d\mathbf{k}}{(2\pi)^3}\sum_\alpha n_\alpha \nu_{\alpha\alpha}(\mathbf{k})\int d\mathbf{v}_\alpha f_\alpha(\mathbf{v}_\alpha) \frac{1-\varepsilon^+(\mathbf{k}\mathbf{v}_\alpha,\mathbf{k})\varepsilon^-(\mathbf{k}\mathbf{v}_\alpha,\mathbf{k})}{\varepsilon^+(\mathbf{k}\mathbf{v}_\alpha,\mathbf{k})\varepsilon^-(\mathbf{k}\mathbf{v}_\alpha,\mathbf{k})}. \tag{П18}$$

С помощью тождества (П8) можно снова привести эту формулу к виду, удобному для рассмотрения предельных случаев (метод вычислений аналогичен вышеизложенному). Окончательный результат таков (здесь использована формула (15)):

$$\Delta P = -\frac{1}{24\pi L^3}\left[\frac{\Theta_e+\mu\Theta_i}{1+\mu} + 2(\Theta_e-\Theta_i)\sum_{n=0}^\infty \text{Re}\,\frac{\Phi_n^{3/2}}{\ln\mu + i\pi(2n+1)}\right],$$

$$\Phi_n = \sum_\alpha \left(\frac{L}{L_\alpha}\right)^2 \int_{-\infty}^{+\infty}\frac{xf_\alpha(x)\,dx}{x-a_n-ib_n}, \tag{П19}$$

$$\left.\begin{array}{c}a_n \\ b_n\end{array}\right\} = \sqrt{\pm\frac{\ln\mu}{u_i^{-2}-u_e^{-2}} + \frac{1}{|u_i^{-2}-u_e^{-2}|}\sqrt{\ln^2\mu+\pi^2(2n+1)^2}}.$$

Из двух значений корня $\sqrt{\Phi_n}$ берется тот, который имеет положительную мнимую часть.

Когда $\Theta_e \gg \Theta_i$, основной вклад в давление дает $g_{ii}(\kappa)$. Из (П19) получается следующий результат: $\quad \Delta P = \dfrac{1}{36\pi^2}\Theta_e \kappa_i^3.$ \hfill (П20)

## ЛИТЕРАТУРА

Moden im Intensitätsspektrum bei genügend grösser zeitlicher Auflösung eine periodische Intensitätsmodulation mit unterschiedlichen Modulationsgraden. Die Modulationsfrequenz stimmt mit dem Frequenzabstand $\Delta f$ der Komponenten (a) und (b) in der Abb. 1 überein und nimmt ebenfalls mit wachsendem $\delta$ bis zum Synchronisationspunkt linear ab. Die Abb. 2 zeigt zwei Interferogramme einer Mode mit Intensitätsmodulation, die nacheinander bei verschiedenen $\delta$-Werten aufgenommen wurden. Ab $\delta = 0.32$ treten im Differenzfrequenzspektrum zwei weitere Komponenten

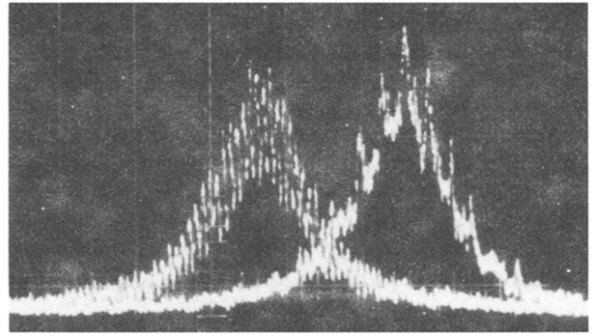

Fig.2. Bei verschiedenen $\delta$ nacheinander aufgenommene Interferogramme einer Laser-Eigenschwingung mit Intensitätsmodulation.

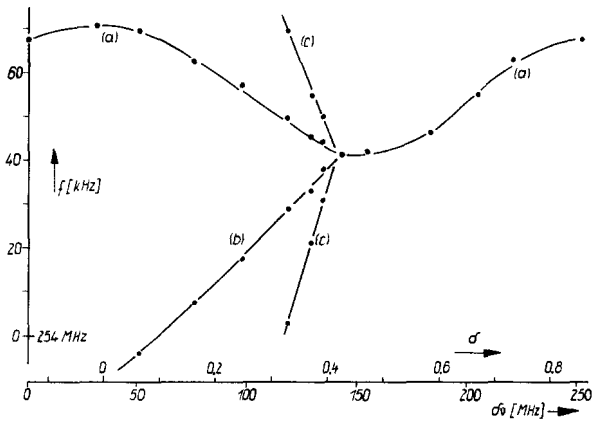

Fig. 1. Darstellung des Zwischenfrequenzspektrums über der relativen Modenverschiebung $\delta$. (Die Stellung $\delta = 0$ wurde aus der Gleichheit der Modenintensitäten von $\nu_2$ und $\nu_3$ bestimmt).

(c) auf, die zu (a) bzw. (b) denselben Frequenzabstand $\Delta f$ haben und als Seitenbänder gedeutet werden können.

Es liegt nahe, diese Erscheinung als eine Energieaustauschwingung anzusehen. Ihre Frequenz nimmt mit Annäherung an den Synchronisationspunkt linear ab. Das Auftreten solcher Koppelschwingungen wird in der Theorie von Lamb [2] als ein Ausdruck nichtlinearer Modenwechselwirkung beschrieben.

*Referenzen*
1. R. A. McFarlane, Phys. Rev. 135 (1964) A 543;
   R. L. Fork and M. A. Pollack, Phys. Rev. 139 (1965) A 1408.
2. W. E. Lamb Jr., Phys. Rev. 134 (1964) A 1429.

\* \* \* \* \*

# SUR LES FONCTIONS CORRELATIVES ET SUR LA PRESSION DE LA PLASMA NON-ISOTHERMIQUE

E. S. MÉDVÉDÉV
*Succurale d'Institut de Physique Chimique. Noginsk. Moscou*

The presence of the "slow" waves in a plasma with two different temperatures give rise to the strong correlations in the positions of the particles. The additional pressure $\Delta P$ due to the Coulomb interaction is shown to be positive ($\Delta P < 0$) et equilibrium).

Il s'agit de la plasma qui se compose de deux sortes des particules dont les densités sont égales ($n_e = n_i \equiv n$). Les températures satisfont soit 1) $\theta_e/\theta_i \gg 1$, soit 2) $\theta_i/\theta_e \gg m_i/m_e$ ($m_i$ et $m_e$ sont les masses des ions et des électrons: $m_i \gg m_e$). Comme on le sait sous ces conditions existent les ondes qui s'affaiblissent legèrement dont les vitesses de phase sont entre les vitesses thermiques des particules: max $(\sqrt{\theta_e/m_e}, \sqrt{\theta_i/m_i}) >$
$> \omega/k > $ min $(\sqrt{\theta_e/m_e}, \sqrt{\theta_i/m_i})$. Nos résultats montrent que cela entraîne les corrélations fortes des positions des particules "lentes" qui agissent à leur tour sur les fonctions thermodynamiques. Notamment la différence $\Delta P = P - n(\theta_e + \theta_i)$ (où $P$ est la pression) qui est négative sous l'équilibre $\theta_e = \theta_i$ change son signe.

Pour calculer les fonctions corrélatives, on utilise les équations de Bogolubov [1]. Les équations correspondantes pour les systèmes quantiques étaient décidées par Syline [2], dont le procédé est entièrement valide pour la plasma classique (on peut prendre une limite $\hbar \to 0$ dans ses formules). Sous $\theta_e/\theta_i \gg 1$ on a:

$$ng_{ii}(k) = -[1+(kL_i)^2]^{-1} + \tfrac{1}{2}[1-\text{sign}(k-k_i)][1+(kL_e)^2] \quad (1)$$

$$ng_{ee}(k) = -\frac{\theta_i}{2\theta_e}[1+(kL_i)^2]^{-1} - \tfrac{1}{2}\text{sign}(k-k_i)[1+kL_e)^2]^{-1} \quad (2)$$

$$ng_{ie}(k) = \frac{\theta_i}{\theta_e}[1+(kL_i)^2]^{-1} + \tfrac{1}{2}[1-\text{sign}(k-k_i)] \quad (3)$$

où

$$k_i L_i = \left(2\left|\ln\frac{m_i \theta_e^3}{m_e \theta_i^3}\right|\right)^{-\tfrac{1}{2}}, \quad L_\alpha^2 = \frac{\theta_\alpha}{4\pi n_\alpha e_\alpha^2} \ (\alpha = i, e),$$

$$\text{sign } x = \frac{x}{|x|}.$$

$\Delta P$ peut se mettre sous la forme:

$$\Delta P = \tfrac{1}{6}\int\frac{d^3k}{(2\pi)^3}\sum_{\alpha,\beta}\frac{4\pi e_\alpha e_\beta}{k^2}n_\alpha n_\beta g_{\alpha\beta}(k). \quad (4)$$

On voit que dans la somme (4) la deuxième terme de $g_{ii}(k)$ joue un rôle dominant. Son apparition est expliquée par l'existance des ondes dont on a parlé précédemment, tandis qu'il est absent sous l'équilibre. Il en résulte:

$$P = \frac{1}{36\pi^2}\theta_e k_i^3. \quad (5)$$

Si on a $\theta_i/\theta_e \gg m_i/m_e$, il faut remplacer les indices $i \rightleftharpoons e$ dans (5).

Il existe un simple moyen pour calculer $g_{\alpha\beta}(k)$ [3]: la méthode d'une particule d'épreuve. Dans le domaine le plus important $L_e^{-1} < k < L_i^{-1}$ il donne un résultat correct mais indéterminé (on n'en sait que $k_i L_i \approx 1$).

*References*
1. N. N. Bogolubov, Les problèmes de la théorie dynamique dans la physique statistique, Moscou (1949).
2. V. P. Syline, Zh. Experim. i Teor. Phys. 40 (1961) 1768.
3. M. N. Rosenbluth and N. Rostoker, Phys. Fluids 5 (1961) 776.

\* \* \* \* \*

## ERRATUM

R. Tsu and F. C. Whitmore, Low temperature breakdown in silicon Zener diode, Physics Letters 24A (1967) 5.

On page 6, column 2,

(1) $3\pi\mu_1 E^2/32c^2$ should be $3\pi\mu_1^2 E^2/32c^2$

(2) $H(x) \equiv \dfrac{\ln[1+(2kT_0\epsilon_r/e^2N_A^{\tfrac{1}{3}})^2 x^4]}{\ln[1+(2kT_0\epsilon_r/e^2N_A^{\tfrac{1}{3}})^2]}$

(3) All $T_e$ or $T$ subsequent to 20th line should be $T_0$.

\* \* \* \* \*